\documentclass[aps,pre,10pt,superscriptaddress,groupedaddress]{revtex4-1}

\usepackage{graphicx}
\usepackage{xcolor} 
\usepackage{dcolumn}
\usepackage{bm}
\usepackage{amssymb}   
\usepackage{amsmath}
\usepackage{subfigure}
\usepackage{epstopdf}
\usepackage{makecell}

\newcommand{\beq}{ \begin{equation} }
\newcommand{\eeq}{ \end{equation} }
\newcommand{\beqs}{ \begin{eqnarray} }
\newcommand{\eeqs}{ \end{eqnarray} }

\usepackage{color}
\usepackage[left]{lineno}

\begin{document}

\title{ Spreading dynamics of a droplet impacting a sphere }

\author{Ming Long}
\affiliation{ \'Ecole Polytechnique, 91128 Palaiseau Cedex, FRANCE }  
\affiliation{Equally contributed}
\author{Jalil Hasanyan}
\affiliation{Radial Research and Development, 
Fairborn, OH, 45324, USA  } 
\affiliation{Equally contributed}
 \author{Sunghwan Jung} \email{sunnyjsh@cornell.edu}
\affiliation{Department of Biological and Enviromental Engineering, 
 Cornell University, Ithaca, NY 14853, USA }  




\date{\today}

\begin{abstract}

In nature, high-speed rain drops often impact and spread on curved surfaces e.g. tree leaves. Although a drop impact on a surface is a traditional topic for industrial applications, drop-impact dynamics on curved surfaces in natural situations are less known about. 
In the present study, we examine the time-dependent spreading dynamics of a drop onto a curved hydrophobic surface.
We also observed that a drop on a curved surface is spreads farther than one on a flat surface. To understand the spreading dynamics, a new analytical model is developed based on volume conservation and temporal energy balance. This model converges to previous models at the early stage and in the final stage of droplet impact. 
We compared the new model with measured spreading lengths on various curved surfaces and impact speeds, which resulted in good agreement.
\end{abstract}

\pacs{Valid PACS appear here}

\maketitle

\section{Introduction}

The impact of drops onto solid surfaces has been investigated for a long time due to its application in industrial processes such as cooling, spray-painting, or ink-jet printing. A similar phenomenon can also be observed in nature as high-speed raindrops impact onto tree leaves \cite{Kim2020,Bhosale2020}.  
{Most leaves are not flat, but curved down \cite{Liu:2010ch}. For biologists, the leaf curvature is an indicator of the leaf's conditions. For example, unhealthy leaves are known to be curved upward or with a high curvature due to a fungus or virus growing on the surface \cite{Anonymous:-caHTDSn,KAWANO:2003io,Curtis:1958ud} or due to the water stress \cite{Fuchs2021,Yuk2022}. When a raindrop hits a leaf surface, the drop experiences impact on a curved surface rather than a flat one. Hence, a droplet impacting a curved surface is important to understand the spreading dynamics. However, spreading dynamics in natural settings are more complicated because most leaves are not only hydrophobic but also elastic and curved.}

{Droplet impact dynamics on a rigid hydrophobic surface have been extensively studied in the context of interfacial motions in simple settings \cite{Yarin:2006en, R2009,Grinspan:2010gt,Roisman:2002ij,Gordillo2019}.}
Roisman \cite{R2009}, developed a model based on assumption on the universal flow and residual thickness of the lamella. Roisman obtained a semiempirical relation for the maximal spreading factor,\linebreak $\frac{R_\mathrm{max}}{R_0} \approx 0.87 \mathrm{Re}^{1/5}-0.40 \mathrm{Re}^{2/5}\mathrm{We}^{-1/2}$ where $R_\mathrm{max}$ is the maximum spreading radius, $\mathrm{We}=\rho U_0^2 (2R_0)/\gamma$ the Weber number, $\mathrm{Re}=\rho U_0 (2R_0)/\mu$ is the Reynolds number, $U_0$ the initial drop velocity, $\rho$ the density of the drop, $\mu$ the dynamic viscosity, $\gamma$ the surface tension. Another proposal was carried out by Clanet, B\'eguin, Richard, and Qu\'er\'e \cite{CBRQ2004}, they proposed that the maximum spreading radius $R_\mathrm{max}$ scales as $R_0^{1/4}$ on partially wettable surfaces using low viscosity liquid (like water). As this relation is not consistent with simple energy balance, an explanation is that kinetic energy is transformed not only to surface energy after impact, but also to internal kinetic energy such as vortical motions. Pasandideh-Fard et al. \cite{PQCM1996} derived another expression for maximum spreading factor, $\xi_\mathrm{max}$ (defined as the maximal spreading length normalized by the initial drop diameter), from the total energy balance which can be calculated as $\xi_\mathrm{max}=\sqrt{\frac{\mathrm{We}+12}{3(1-\cos \theta)+4\frac{\mathrm{We}}{\sqrt{\mathrm{Re}}}}} $ where $\theta$ is the contact angle.

Rioboo, Marengo and Tropea \cite{RMT2002}, categorized the time evolution of the spreading factor into four phases: kinematic, spreading, relaxation, and wetting/equilibrium phases. In the kinematic phase ($t^{*} < 0.1$ where $t^*=tU_0/2R_0$ is dimensionless time), the spreading factor grows corresponding to a power law; in which the experimental exponent lies between 0.45 and 0.57. Other studies such as Bird et al. \cite{BTS2009} and Kim et al. \cite{KFC200} seem to support a similar exponent in accordance with experimental data (As Kim et al. point out [6], little data of early spread exists because of the difficulty of measurement). Roisman, Berberovi and Tropea \cite{RBT2009} also developed a model for a spreading thin free liquid sheet. They identified three regimes in the time dependence of the height of the deforming drop at the symmetry center. During the first two main regimes, the dimensionless central height as a function of the dimensionless time ($h_c^*=h_c/2R_0$) can be approximated as follows,  $h_c^{*} \approx 1-t^{*}$ at $ t^{*} < 0.4$, and $h_c^{*} \propto 1/t^{*2}$ at $0.4 < t^* < t_\mathrm{viscous}$, in which $t_\mathrm{viscous}$ is large and difficult to determine precisely.

The impact of a droplet on a curved surface has also been investigated recently. Hung and Yao \cite{HY1999} studied the impact of micrometric droplets on cylindrical wires. Droplets impacting wires either disintegrate or drip depending on the Weber number and the Bond number. Others studied the aftermaths of the impact of liquid drops on cylindrical surfaces \cite{LGM2014,Arogeti2019}, and rebound, coalescence or disintegration phenomena in both experimental or computational approaches \cite{Liu:2010ch,Banitabaei2017,Khojasteh2017,Chen2017}. In addition, several studies investigated the impact of a droplet on a spherical target \cite{BRT2007,Chen2017,Banitabaei2017,Iwamatsu2018,Khojasteh2019,Liu2019}.  

In this paper, we study the effect of curvature on the spreading of a water drop onto a hydrophobic sphere and develop models of the phenomenon.


\section{Experimental Procedure}

\begin{figure}
 \centering
 \includegraphics[width=0.7\textwidth]{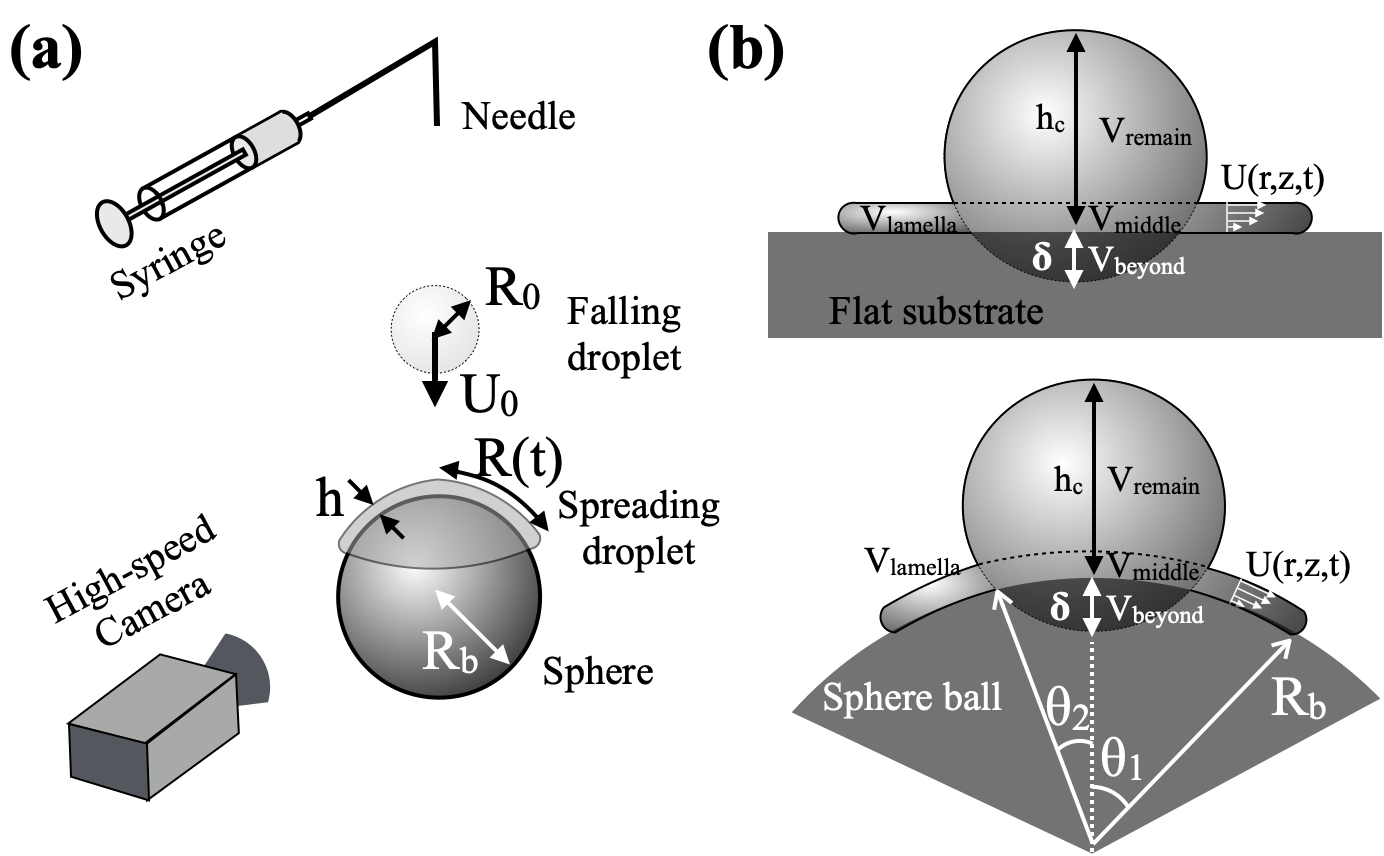}
 \caption{(a) Schematic diagram of experimental setup. (b) Schematics of different volumes of a droplet on a flat surface (upper panel) or on a curved surface (lower panel). }
 \label{ExpSetup}
\end{figure}

The experimental setup is illustrated in Fig. \ref{ExpSetup}a. The curved surfaces in the experiments were white Delrin spheres (McMaster Co.) of four different sizes: 1.27, 1.90, 2.54, and 3.81 cm in diameter. The spheres were uniformly coated with a hydrophobic spray (WX2100, Cytonix Co.). As a result, the contact angles of the samples were increased from 50-60 to 120-130 degrees.
For experiments on a flat surface, a plexiglass plate was coated with the hydrophobic spray. 

Drops were generated using a needle with a 0.34mm inner diameter (23 gauge). The vertical position of the needle was changed from 5 to 50 cm to vary the impact velocity. In addition, the horizontal position of the needle was controlled with an XY stage (Thorlabs Co.) to precisely release the droplet at the apex of the balls.
Two different high-speed cameras were used depending on the frame rate needed: Photron Fastcam-APX RS at either 6000 or 20 000 fps, and IDT N3 at 1000 fps. 

From experiments, we measured five major quantities: the initial drop size $R_0$ and speed $U_0$, the time evolution of the spreading radius $R(t)$, the maximal spreading radius $R_\mathrm{max}$, and the central height of the impacting drop $h_c(t)$ Figure \ref{ExpSetup}. These measurements are performed either automatically using Matlab, or manually using ImageJ when images can be difficult to analyze automatically.
\begin{figure}
 \centering
 \includegraphics[width=0.65\textwidth]{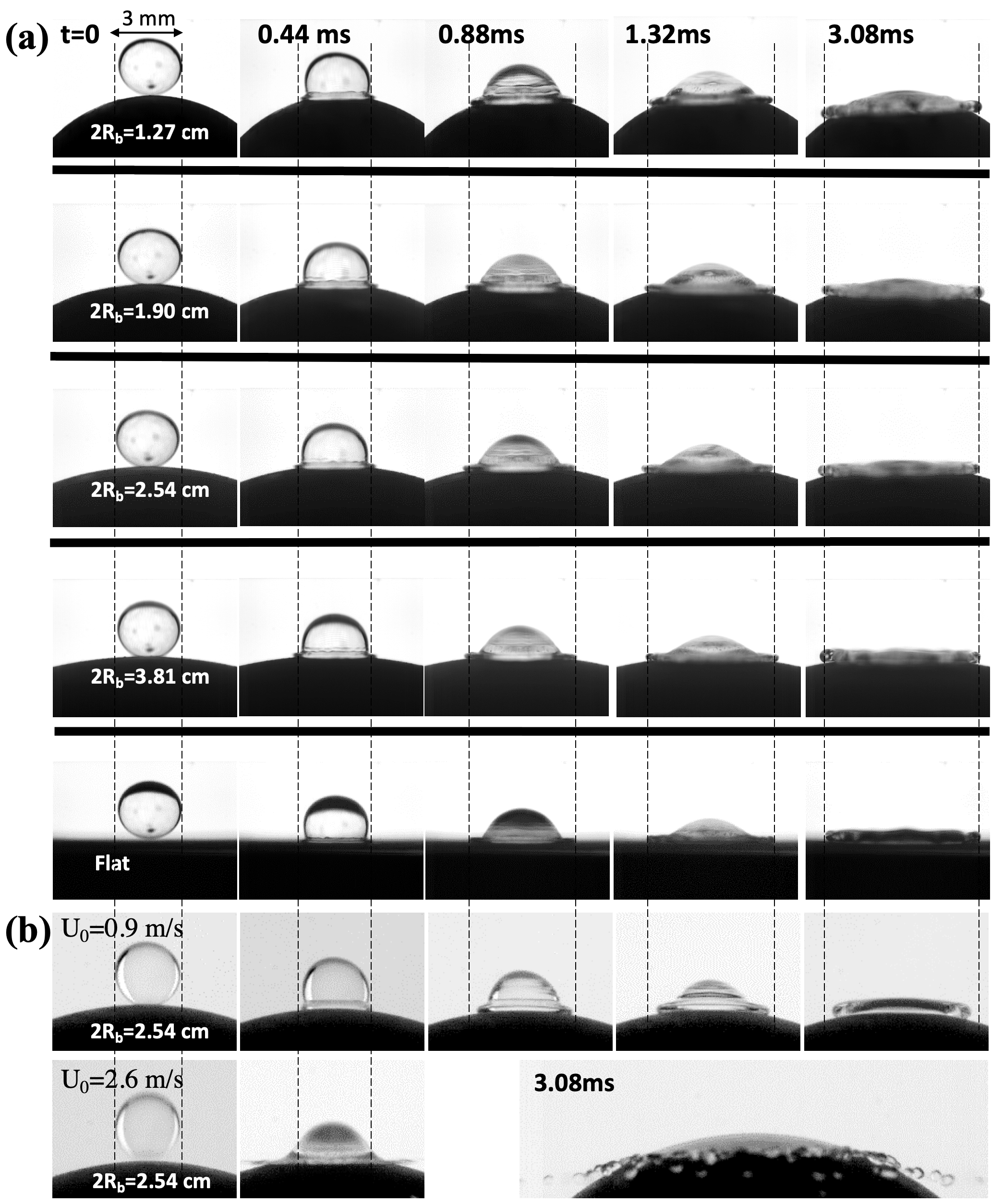} 
 \caption{(a) Image sequences of a drop impacting on a sphere of $2R_b=$ 1.27, 1.90, 2.54 and 3.81 cm in diameter and a flat surface. The impact speed is at $U_0=1.2$ m/s, and the drop is $2R_0=3$ mm. (b) Image sequences of a drop impacting on a sphere of 2.54 cm in diameter at different speeds ($U_0 = 0.9$ and 2.6 m/s). }
 \label{ImSeq}
\end{figure}


\section{Theoretical model}

\subsection{Volume conservation}

{Initially, a falling drop is considered as a sphere of radius $R_0$ falling with a velocity $U_0$ as shown in Fig. \ref{ExpSetup}a. However, at the moment of impact, the geometry of a spreading drop has a segment of a spherical drop and a thin lamella on its surface. As shown in Fig. \ref{ExpSetup}b, we can divide the impacting drop into three distinct volumes: (i) the volume of the remaining part of the original impacting drop ($V_\mathrm{remain}$), (ii) the volume in contact with the solid surface ($V_\mathrm{middle}$) and (iii) the volume of the lamella ($V_\mathrm{lamella}$). The volume conservation which is a sum of these three volumes ($V_\mathrm{remain}+V_\mathrm{middle}+V_\mathrm{lamella}$) equals to its orignial volume ($\frac{4\pi}{3} R_0^3$) can be expressed in terms of a spreading radius $R(t)$, a lamella thickness $h(t)$, and other geometric coordinates and lengths. This volume-conservation equation becomes
\beq
(R_b+R_0-\delta) [ 8(1-\cos \theta_1)h(3R_b^2 +3R_b h +h^2) -(h+\delta)^2 (3R_b+3R_0+2h-\delta)] +3 (\delta+h)^2(R_0-R_b-h)^2=0 \label{eq_mass} \,, 
\eeq
which is solved further with an energy conservation equation.
}

\subsection{Energy balance}

{
Total energy is assumed to be constant at every moment. Therefore, a sum of the initial kinetic energy ${\cal E}_{k}^{(initial)}(=\rho  (2/3) \pi R_0^3 U_0^2)$ and the surface energy ${\cal E}_{s}^{(initial)}(=\gamma 4 \pi R_0^2)$  should be equal to the sum of the subsequent kinetic energy ${\cal E}_{k}(t)$, surface energy ${\cal E}_{s}(t)$, and viscous work $W$ as follows
\beq
{\cal E}_{k}^\mathrm{(initial)}+{\cal E}_{s}^\mathrm{(initial)}={\cal E}_{k}(t)+{\cal E}_{s}(t)+\int_0^t W dt' \,. \label{Energy}
\eeq
Instantaneous kinetic energy ${\cal E}_k(t)$ can be estimated by adding the kinetic energy of the remaining volume of the original drop and the kinetic energy of the lamella. The kinetic energy of the lamella can be calculated using an assumption of the semiparabolic velocity in the lamella. This semiparabolic velocity profile is often assumed for a free-surface flow \cite{Batchelor:1967wd,Landau:2012ws,Lamb:1945wl}. }

\begin{figure}
\centering
 \includegraphics[width=0.45\textwidth]{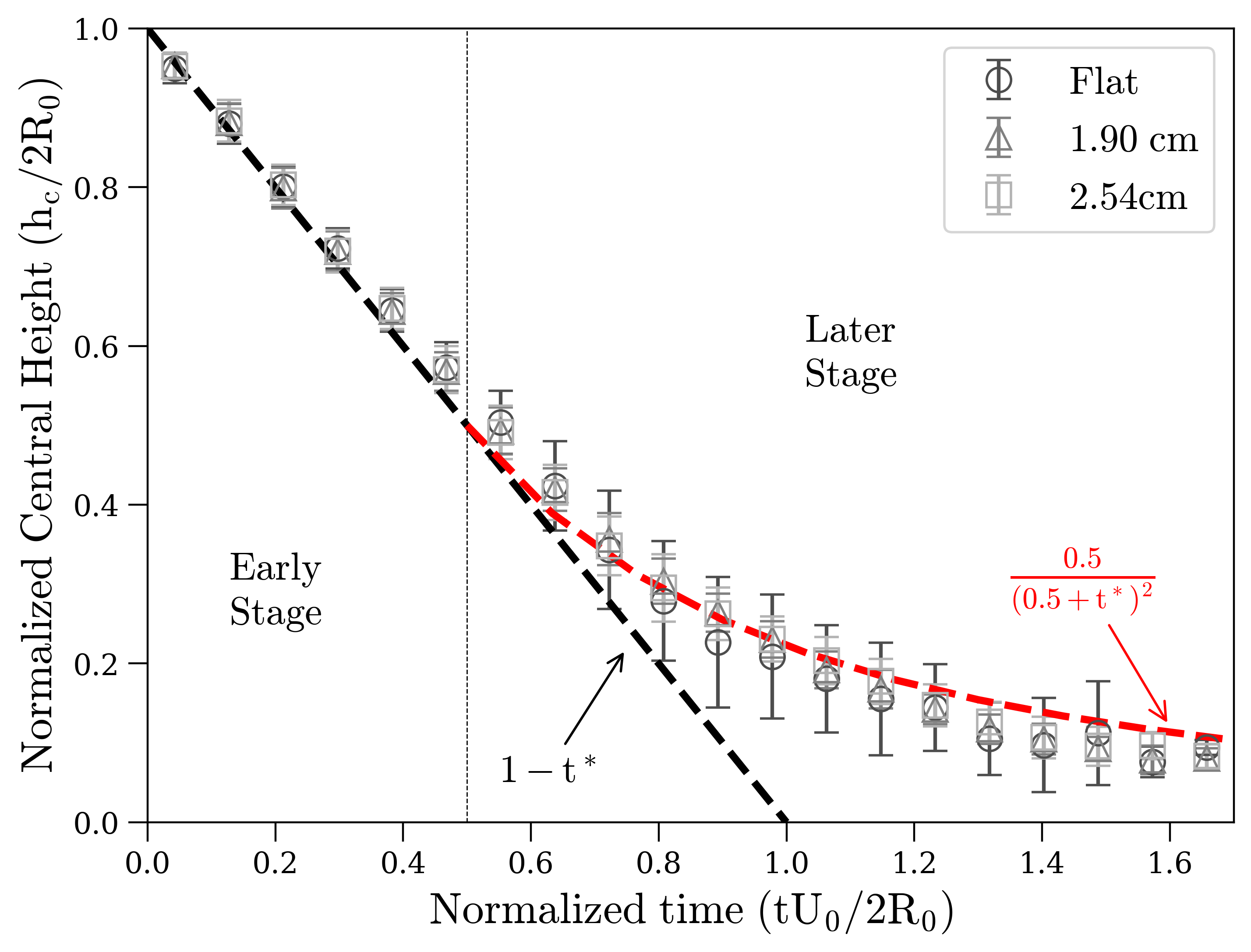}
 \caption{Measurements of the time-evolution of normalized central height ($h_c/2R_{0}$) with normalized time ($tU_{0}/2R_{0}$). Symbols represent experimental data with different spheres. Experimental data shows $h_c/(2R_0) = 1- t^*$ in the beginning ($t^* < 0.5$) and then follows $h_c/(2R_0) = 0.5/(0.5+ t^*)^2$ later.  }
 \label{hc_time}
\end{figure}

{Next, surface energy ${\cal E}_{s}(t)$ depends on the area of the water-air and water-solid surfaces as ${\cal E}_{s}(t)=\gamma A_{LG}(t) + (\gamma_{LS}-\gamma_{SG})A_{LS}(t)$. The subscripts for surface area ($A$), $L$, $G$, and $S$ represent liquid, gas, and solid, respectively. 
The two areas ($A_{LG}(t)$, $A_{LS}(t)$) will be expressed in terms of $R(t)$, $h(t)$, and other known geometric parameters. Therefore, the total energy balance becomes 
\beqs
&&  \gamma \left[2\pi (R_b+h)^2(\cos\theta_2-\cos\theta_1) + 4\pi R_0^2-2\pi R_0\delta + 2\pi \left( R_b+ \dfrac{h}{2} \right) \sin\theta_1 h -2\pi R_b^2\cos\theta (1-\cos\theta_1)\right]  \nonumber \\
&-& 4\gamma \pi R_0^2  + \pi \rho\dot{h}_c^2 \left[ \dfrac{2 R_{0}^3}{3} -\dfrac{V_\mathrm{beyond}}{2\pi} + \dfrac{3 s_0^4}{10 h} \ln\left(\dfrac{\theta_1}{\theta_2}\right)\right] -\dfrac{2\pi \rho U_0^2 R_0^3 }{3} + W=0 \label{eq_energy} \,.
\eeqs 
}

{For energy dissipation due to the viscous work ($W$), we consider only shear rates along lamella as $W = \int_{V_\mathrm{lamella}} \mu ({\partial U}/{\partial z})^2 \mathrm{d} \Omega $
where $\mu$ is the fluid viscosity and $\Omega$ is the lamella volume. 
Here, the lamella velocity ($U$) is assumed to be semiparabolic in the lamella, which incorporates zero stress on the free surface. Therefore, on the spherical surface, the radial velocity is expressed as $U(s,z,t)=\overline{U}(t)\frac{R_b \theta_2}{s}\frac{z(2h-z)}{h^2}$ where $z$ is the coordinate normal to the surface, $s$ is the curvilinear abscissa that can be easily calculated as $s=R_b\theta(t)$. Finally, the velocity of the lamella is given as 
\beq
U(s,z,t) = \frac{3 |\dot{h}_c(t)|s_0^2(t) }{4h^3(t)s}z(2h-z) \, ,
\eeq
where $s_0(t)=R_b \arcsin({a}/{R_b})$. Here $a$ is the half-length of $V_\mathrm{beyond}$ (see the full expression in Equation \ref{curved_curvilinear_abscissa} of Appendix B). By taking the $z$-gradient on the above velocity profile, the energy dissipation ($W$) is calculated. }

\begin{figure}
 \centering
 \includegraphics[width=0.45\textwidth]{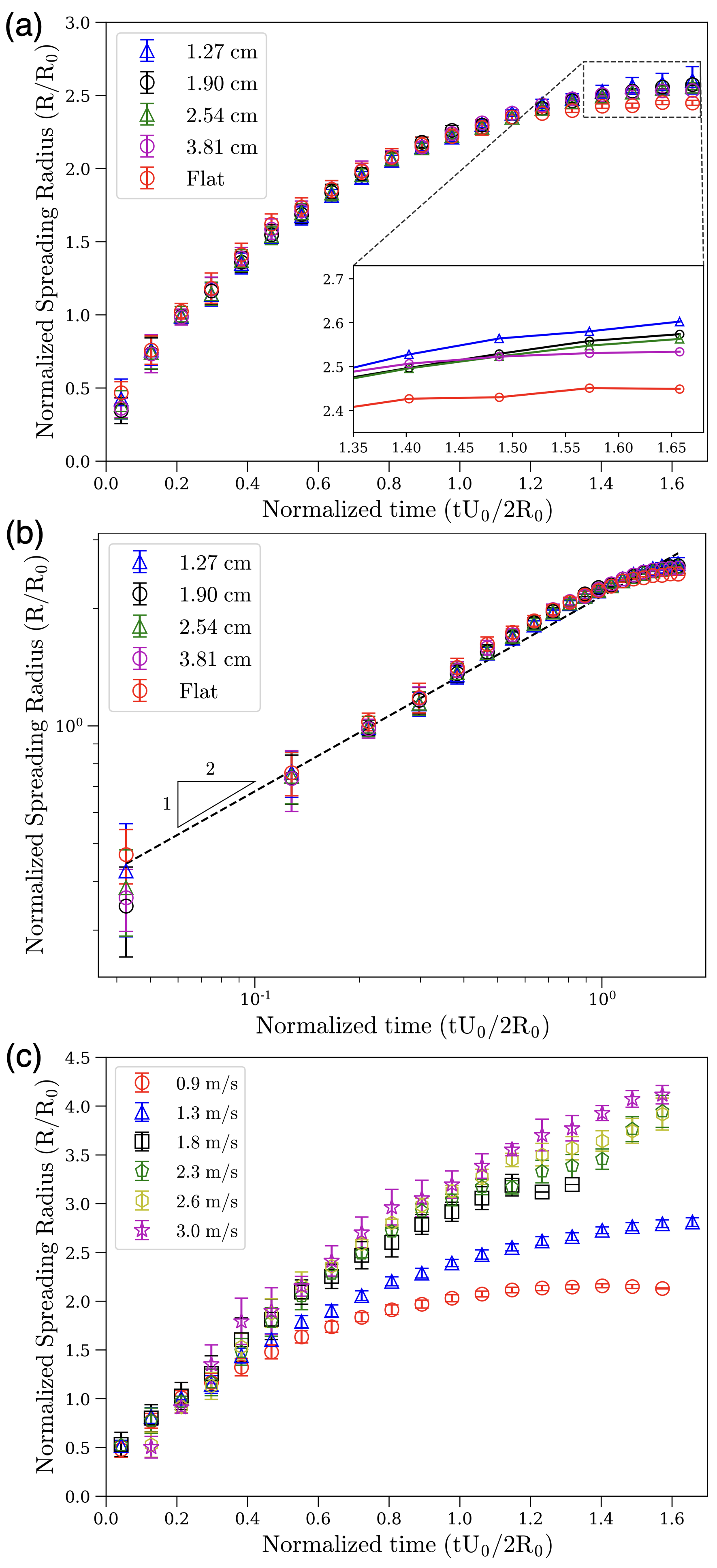}
 \caption{ Measurements of normalized spreading radius ($R(t)/R_{0}$) with normalized time ($tU_{0}/2R_{0}$) (a,b) with different radii of curvature at a speed of 1.2 m/s and (c) with different velocities for a sphere with 2.54 cm diameter. In the panel (b), the dotted line is $R(t)/R_0 = 2.15 (tU_{0}/2R_{0})^{1/2}$ to confirm a power-law increase with an exponent of 1/2. }
 \label{comparison_Rbcst}
\end{figure}

\subsection{Numerical and asymptotic solutions}

{We have two sets of equations; one from volume conservation (Eq. \ref{eq_mass}) and the other from energy conservation (Eq. \ref{eq_energy}). There are two unknowns to be determined; lamella thickness $h(t)$ and spreading radius $R(t)$. 
This set of equations can be numerically solved to determine the two unknowns. Hence, at every moment, an appropriate region in the $(R,h)$ space was meshed thinner and thinner according to the expected set of $(R,h)$ from the previous iteration. Both equations were computed at every node, and the solutions were then calculated as the coordinates of the minimum value of the multiplication of both equations (Eqs. \ref{eq_mass} and \ref{eq_energy}). }

{First, let's check two asymptotic solutions analytically. 
At the early stage ($t \ll 1$), we can assume both spreading radius and penetration depth $R(t), \delta \approx U_0 t \ll R_0$, $R_b$, so that $\sin \theta_1 \approx R(t)/R_b$ and the viscous work on the lamella can be considered as negligible. Then, simplifying volume conservation and energy balance yield at early stage 
\beq 
R(t) \approx R_0\left( \frac{4R_b}{R_b+R_0} \right)^{1/2} \left( \frac{U_{0}}{2R_{0}} t \right)^{1/2}  \,. \label{eq_spread}
\eeq

{At the later stage ($t\gg 1$), we can assume the penetration depth $\delta \approx 2R_0$ and then $\theta_2 \ll 1$, $h \ll R_b$, so that the total energy balance combined with the volume conservation can be simplified as
\beq
\mathrm{We} \left[ 1-\frac{4}{\sqrt{\mathrm{Re}}} \frac{R(t)^2}{R_0^2} \right]=6 \frac{R_b^2}{R_0^2} (1-\cos \theta_{\mathrm{Advance}})(1-\cos \theta_1) - 12 \,. \label{sp_dyn_curve}
\eeq
In terms of the maximum spreading radius, we approximate $R(t)$ and $\theta_1$ as $R_\mathrm{max}$ and $R_\mathrm{max}/R_b$, respectively. Then, we can solve for $R_\mathrm{max}$ at given parameters ($\mathrm{We}, \mathrm{Re}, R_0, R_b$, and $\theta_{\mathrm{Advance}}$).

In the limit of the flat case ($R_b \rightarrow +\infty$, $1-\cos \theta_1 \approx \frac{1}{2}(R^2(t)/R_b^2)$, $R(t\rightarrow +\infty) = R_\mathrm{max}$), the above spreading equation becomes 
\begin{equation}
\frac{R_\mathrm{max}}{R_0}=\sqrt{\frac{\mathrm{We}+12}{3(1-\cos \theta_{\mathrm{Advance}})+4 {\mathrm{We}}{({\mathrm{Re}}})^{-1/2}}} \,.
\end{equation}
This drop-spreading equation is exactly same as the one from Pasandideh-Fard et al. \cite{PQCM1996}. 
It is noteworthy that a drop will spread less on a hydrophobic surface ($\theta_{\mathrm{Advance}} \rightarrow \pi$).
}

{Like other studies, we can express it in terms of the maximum spreading factor; $\beta_\mathrm{max} \equiv R_b\theta_1 (t \rightarrow +\infty)/R_0$. For larger spheres compared to the drop size ($R_b/R_0 \ll 1$) and smaller spreading distance ($\theta_1 \ll 1$), Eq. \ref{sp_dyn_curve} becomes
\beqs
&& \mathrm{We} \left[ 1-\frac{4}{\sqrt{\mathrm{Re}}}  \beta_\mathrm{max}^2 \right]=3  (1-\cos \theta_{\mathrm{Advance}}) \beta_\mathrm{max}^2 - 12 \nonumber  \\ 
&\Rightarrow& \beta_\mathrm{max}= \sqrt{\frac{\mathrm{We}+12}{3(1-\cos \theta_{\mathrm{Advance}})+4 {\mathrm{We}}{({\mathrm{Re}}})^{-1/2}}} \,. \label{sp_dyn_curve2}
\eeqs
It is worth noting that this is an approximate solution with several assumptions listed above. Also, this is different from the previous analytical model in \cite{Liu2019} ($8/\beta_\mathrm{max}$ is missing). 
}


\begin{figure}
 \centering
 \includegraphics[width=0.45\textwidth]{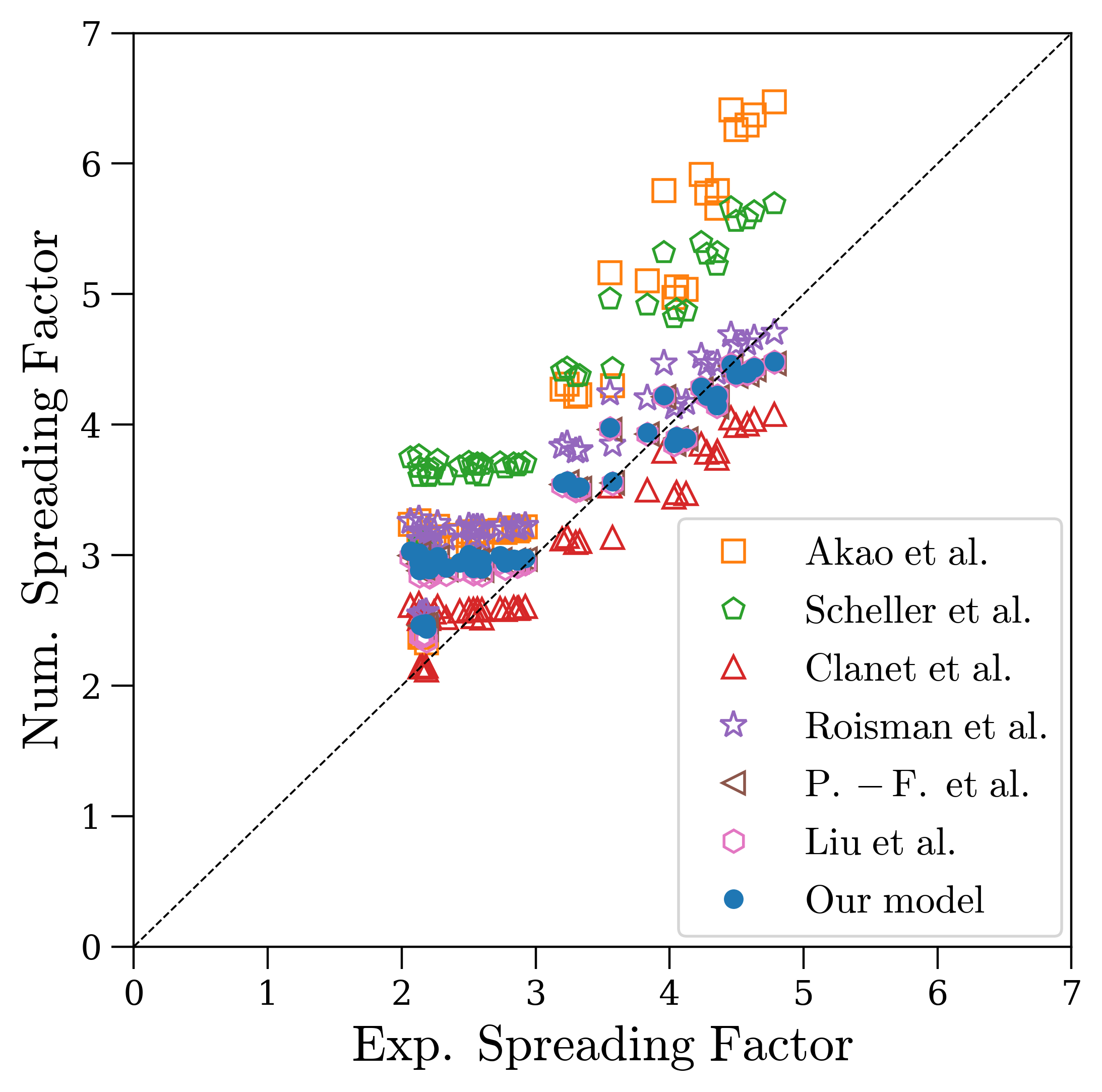}
 \caption{ Comparison of experimental spreading factor $R_\mathrm{max}/R_0$ with theoretical normalized maximal spreading factor from various studies. Please refer to Table \ref{model_tab} for detailed expressions.   }
 \label{comparison}
\end{figure}

\subsection{Comparison with experiments}


First, we measured the central height $h_c$ of the impacting drop as in Fig. \ref{hc_time}. At an early stage ($t^*<0.5$), $h_c^*$ decreases linearly with normalized time as $h_c^{*}(t) \simeq 1-t^*$. At the later stage ($0.5 < t^* $), $h_c^*$ deviates from the linear trend and follows $h_c^*(t) \approx 0.5 / (t^*+0.5)^2$. It is also noticeable that this trend remains the same regardless of the curvature of the substrate (see Fig. \ref{hc_time}). 

Next, the effect of curvature on the spreading dynamics is characterized as shown in Fig. \ref{ImSeq}a. A water drop ($R_0=1.5$ mm) impacts spheres of different curvatures at a speed of $U_0=1.2$ m/s. The drop appears to spread almost the same on the projected distance, but has a slightly more spreading distance along the curved surface as the radius of the surface decreases. Figure \ref{comparison_Rbcst}a shows the normalized spreading distance $R^* \equiv R/R_0 = R_b \theta_1/R_0$ versus the normalized time, showing a very slight increase with curvature. The normalized spreading radius $R^*$ initially increases as a power law with respect to normalized time $t^*$ such as $R(t)/R_0 = 2.15 (tU_0/2R_0)^{1/2}$ (see Fig. \ref{comparison_Rbcst}b). This result is consistent with the asymptotic solution derived in the previous section (Eq. \ref{eq_spread}). It is worthy noting that the prefactor value, 2.15, of the best fit is very close to the one from our calculation; $(4R_b/(R_b+R_0))^{1/2} \approx 1.8-1.9$.  Figure \ref{ImSeq}b shows the effect of the impact speed.  The higher the velocity, the longer the maximum spreading radius and the faster it expands (see Fig. \ref{comparison_Rbcst}c).

Finally, Fig. \ref{comparison} shows the comparison of our experiments with various theoretical predictions. All models are listed in Table \ref{model_tab} and the root mean square error (RMSE) is computed for each model as
\begin{align*}
RMSE = \sqrt{\frac{1}{N_\mathrm{exp}}\sum_{i=1}^{N_\mathrm{exp}}{(\beta_\mathrm{max}^\mathrm{num} - \beta_\mathrm{max}^\mathrm{exp})^2}} \,,
\end{align*}
where $N_\mathrm{exp}$ is the number of experiments, $\beta_\mathrm{max}^\mathrm{num}$ and $\beta_\mathrm{max}^\mathrm{exp}$ are the numerical and experiment spreading factors, respectively. Here, we will consider two different regimes: $\beta_\mathrm{max}^\mathrm{exp} \leq 3 $ and $\beta_\mathrm{max}^\mathrm{exp} > 3 $. 
First, Akao et al. \cite{Akao1980} and Scheller et al. \cite{Scheller1995} models seem to systematically overestimate the maximal spreading diameter for all $\beta_\mathrm{max}^\mathrm{exp}$. For small spreading factors $\beta_\mathrm{max}^\mathrm{exp} < 3$, the model of Clanet et al. \cite{CBRQ2004} gives the best agreement with our experimental data. For $\beta_\mathrm{max}^\mathrm{exp} > 3$,  our model, the model of Pasandideh-Fard et al. \cite{PQCM1996} and the model of Liu et al. \cite{Liu2019} show the best results. A higher $\beta_\mathrm{max}^\mathrm{exp}$ is of our interest as the lamella spreads along the surface more than the drop size.

\begin{table}[]
    \centering
    \begin{tabular}{c|c|c|c|c}
        Reference & Predicted model & Type of surface & RMSE  & RMSE  \\
         &  &  & ($\beta_\mathrm{max}^\mathrm{exp} \leq 3 $) &  ($\beta_\mathrm{max}^\mathrm{exp} > 3) $ \\
        \hline
         \rule{0pt}{12pt}Akao et al. \cite{Akao1980} & $0.613 \mathrm{We}^{0.39}$ & Flat & 0.67 & 1.40 \\
         \rule{0pt}{12pt}Scheller et al. \cite{Scheller1995} & $0.61 \mathrm{Re}^{1/5} (\mathrm{We} \mathrm{Re}^{-2/5} )^{1/6}$ & Flat & 1.19 & 1.05 \\
         \rule{0pt}{24pt}Clanet et al. \cite{CBRQ2004} & $ 0.9 \mathrm{We}^{1/4}$ & \makecell{Flat \\ (A 0.9 coefficient is used \\ for fitting in ref. \cite{CBRQ2004})} & 0.25 & 0.46 \\
         \rule{0pt}{12pt}Roisman et al. \cite{R2009} & $0.87 \mathrm{Re}^{1/5} - 0.4 \mathrm{Re}^{2/5} \mathrm{We}^{-1/2}$ & Flat & 0.71 & 0.35\\
         \rule{0pt}{15pt}P.-F. et al. \cite{PQCM1996} & $\sqrt{\frac{\mathrm{We}+12}{3(1-\cos \theta_{\mathrm{Advance}})+4 {\mathrm{We}}{({\mathrm{Re}}})^{-1/2}}}$ & Flat & 0.50 & 0.21 \\
         \rule{0pt}{15pt}Liu et al. \cite{Liu2019} & $\mathrm{We} + 12 + B_g = S_1 + S_2 +4 \frac{\mathrm{We}}{\sqrt{\mathrm{Re}}}\beta_{max}^2$ & Flat and Curved & 0.47 & 0.21 \\
         \rule{0pt}{12pt}Our model & Eq. \ref{sp_dyn_curve} & Flat and Curved & 0.51 & 0.21
    \end{tabular}
    \caption{Various models for predicting the maximum spreading factor.}
    \label{model_tab}
\end{table}

\section{Conclusion and discussion}

{We studied how a droplet impacts and spreads over a sphere, which is inspired by natural phenomena of raindrops impacting on curved biological surfaces. In this work, the spreading dynamics on a sphere has been theoretically formulated and compared with experiments. The underlying concept of this study is to consider a spherical segment and a skirting lamella of a spreading droplet after impact.  First, we characterized the height of the spherical segmanet, $h_c$, which decreases linearly in time first and then decays in the power law. Secondly, by considering the mass balance and the energy balance, we can predict the spreading lamella over time. In the beginning, the lamella increases its radius as $t^{1/2}$, which is verified in experiments. Finally, the maximum spreading factor is theoretically predicted and compared with experimental observations.  }

This study focused only on the drop spreading on a fixed sphere. However, actual leaves will undergo bending and twisting motions \cite{Bhosale2020} and deformations, thereby changing the local surface curvature and angle. It would be interesting to study how a drop spreads on a thin elastic substrate as the bottom surface deforms due to the pressure from the impacting drop. Additionally, the bottom curvature and particles \cite{Esmaili2021} could change the generation rate of drop splashes/aerosols. As the thin lamella moves along a convex surface, it becomes unstable and may generate and eject more splashes. Such splashes could affect both water retenion on plants \cite{Kang2018,Lenz2022} and spore dispersal \cite{Kim2019}, which leads to interesting questions and implications. 

\newpage

\section{Appendix A: Case of a flat rigid surface}
This section will show calculations on a flat surface based on the geometry of an impacting drop, volume conservation and energy balance. 

\subsubsection{Geometry of the model}
{Prior to impact, the droplet assumes a spherical shape with radius $R_0$ while falling at the velocity $U_0$. The part of the liquid in immediate contact with the surface is transferred to the lamella during the impact, i.e., $V_\mathrm{beyond}$ is transferred to $V_\mathrm{lamella}$ in Fig. \ref{ExpSetup}b. The volumes $V_\mathrm{middle}$ and $V_\mathrm{beyond}$ are defined as the volumes of a two-base spherical segment of height $h$, and a spherical cap of depth $\delta$, respectively. The droplet at the point of maximal spreading is assumed to have deformed into a cylinder with radius $R$ and height $h$. }

\subsubsection{Flow in the lamella}
A general form of the radial velocity given by mass conservation in the lamella is assumed to be  $U(r,z,t) = \overline{U} (t)\frac{r_0(t)}{r} \frac{z(2h-z)}{h^2}$ at time $t$, assuming a semi-parabolic profile due to viscosity as shown in Fig. \ref{semiparabola}, and $h$ small so the edge of $V_\mathrm{middle}$ can be approximated by a cylinder of height $h$ and radius $r_0(t)=R_0 \sqrt{1-(1-\delta(t)/R_0)^2}$.

To determine $\overline{U}(t)$, one can apply mass conservation to $V_\mathrm{middle}$, still approximated as a cylinder. Mass coming in through the top surface is assumed to have the same velocity as the apex of the impacting drop $\dot{h}_c(t)$, whose expression was determined experimentally (as shown in the later section). Thus, it can be written as $M_\mathrm{in}=\rho \pi r_0(t)^2 |\dot{h}_c(t)|$. Mass coming out through the edges can be expressed as $M_\mathrm{out}=\rho 2 \pi r_0(t) h (t) U^*(t)$ where $U^*=\frac{1}{h} \int_{0}^{h}U \mathrm{d}z=\frac{ 2 \overline{U} (t)}{3}$ is the mean velocity on the edge of $V_\mathrm{middle}$. Mass conservation of the middle part then yields $\overline{U}(t)=\frac{3r_0(t) |\dot{h}_c(t)|}{4 h(t)}$ and radial velocity 
\begin{equation} 
\label{flat_velocity_profile}
U(r,z,t) = \frac{ 3|\dot{h}_c(t)|r_0^2(t) }{4h^3(t)r}z(2h-z).
\end{equation}

\subsubsection{Governing equations}
The volume conservation is obtained by equating the virtual volume $V_\mathrm{beyond}$ with the lamella volume $V_\mathrm{lamella}$. Using the aforementioned geometry of the flat impact, the volumes in Fig. \ref{ExpSetup}b are given by
\begin{eqnarray}
V_\mathrm{beyond} &=&  \frac{\pi}{3} \delta^2 (3R_0-\delta), \\
V_\mathrm{middle} &=&  \pi h \left(-\frac{h^2}{3}-\delta^2+2\delta R_0 + R_0h-\delta h \right), \\
V_\mathrm{lamella} &=&  \pi R^2 h - V_\mathrm{middle}.
\end{eqnarray}
The volume conservation condition with $V_{beyond}=V_{lamella}$ reduces to
\begin{equation} 
\label{flat_volume_conservation}
3R_0(\delta + h)^2 - 3hR^2 - (\delta + h)^3 = 0.
\end{equation}

 \begin{figure}
 \centering
 \includegraphics[width=0.3\textwidth]{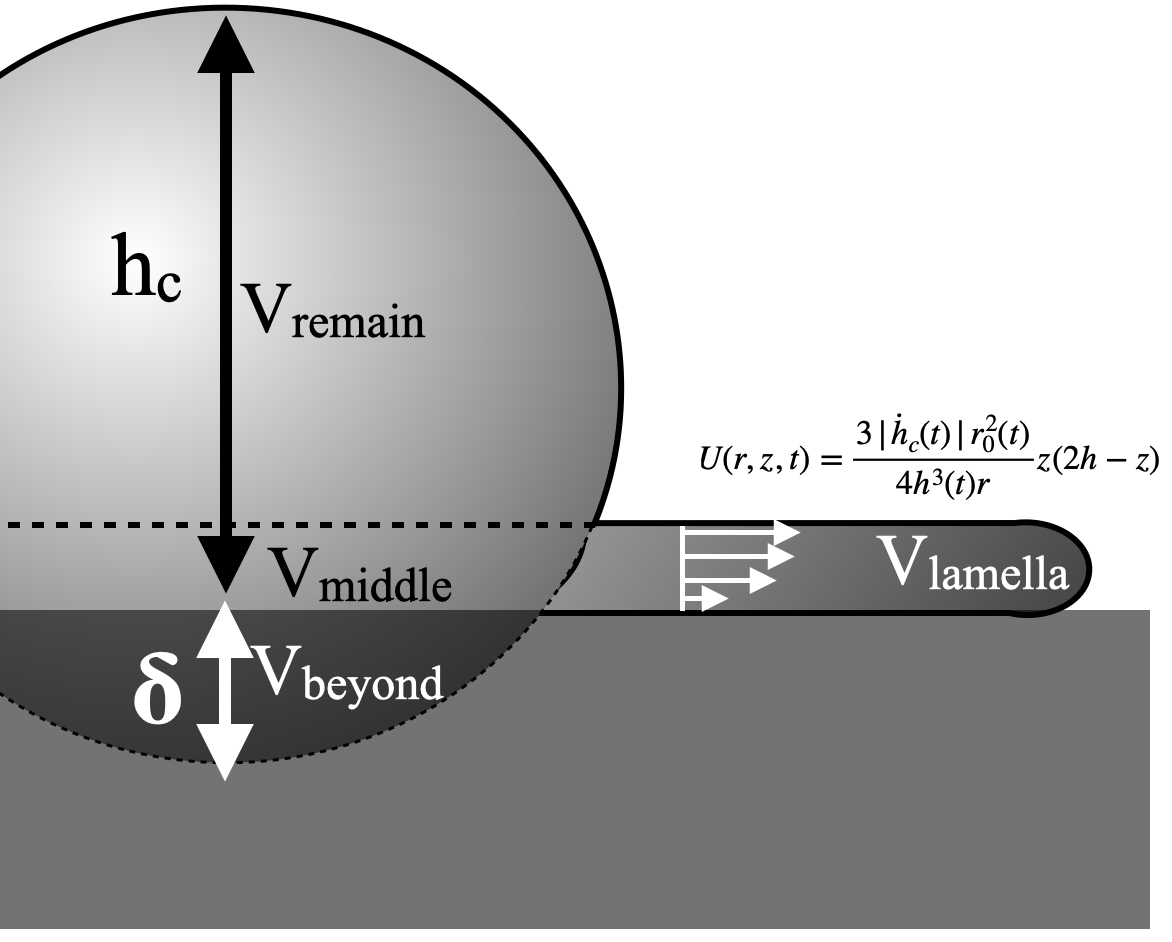}
 \caption{Semi-parabolic velocity profile in the lamella }
  \label{semiparabola}
 \end{figure}

Total energy is assumed to be constant at every moment and described by $E_\mathrm{k_i}+E_\mathrm{s_i}=E_\mathrm{k}(t)+E_\mathrm{s}(t)+W$, where $E_\mathrm{s_i}=\gamma 4 \pi R_0^2$ is the surface energy and $E_\mathrm{k_i}=(1/2) \rho U_0^2 (4/3) \pi R_0^3$ is the kinetic energy. The surface energy over during the impact is 
\begin{equation}
\label{surface_engery}
    E_\mathrm{s}(t)=\gamma S_{LG} + (\gamma_{LS}-\gamma_{SG})S_{LS},
\end{equation}
where the surface subscripts L, G, and S represent liquid, gas, and solid, respectively. The surface areas of liquid-solid and liquid-gas interfaces are
\begin{eqnarray}
\label{flat_surface_energy}
S_\mathrm{LS} &=& \pi R^2, \\
S_\mathrm{LG} &=& S_\mathrm{lamella}+S_\mathrm{remain},  \\
S_\mathrm{lamella} &=& 2\pi R h + \pi R^2 - \pi [R_0^2-(R_0-\delta-h)^2],\\
S_\mathrm{remain} &=& 4 \pi R_0^2 - 2 \pi (\delta+h)R_0.
\end{eqnarray}
The lamella's surface area, $S_{lamella}$, is computed by taking the difference of the surface area of a bottomless cylinder with radius $R$ and height $h$, with the area of a circle with radius $\sqrt{R_0^2-(R_0-\delta-h)^2}$ (i.e., the top surface of $V_{middle}$). The droplet's remaining surface area, $S_{remain}$, is the difference between the surface area of the original droplet and the spherical cap with radius $R_0$ and height $\delta + h$.

The kinetic energy $E_\mathrm{k}(t)$ can be estimated by adding the kinetic energy of remaining portion of the droplet with the kinetic energy of the lamella. Integrating the kinetic energy of the lamella, divided in annuli of volume $2 \pi r dr$, yields $E_k(t) = \frac{1}{2} \rho \dot{h}_c^2 (\frac{4}{3} \pi R_0^3 - V_\mathrm{beyond})+\int_{z=0}^{h} \int_{r_0(t)}^{R(t)} \frac{1}{2} \rho U^{2} 2\pi r \mathrm{d}r \mathrm{d}z$. Using the radial velocity from Eq. \ref{flat_velocity_profile} and the defined volumes, the kinetic energy simplifies to

\begin{equation}
\label{flat_kinetic_energy}
E_k (t)=\pi \rho\dot{h}_c^2 \left[ \frac{2}{3} R_0^3 - \frac{\delta^2}{6} (3R_0-\delta)+\frac{3r_0^4}{10h} \ln \left( \frac{R}{r_0} \right) \right].
\end{equation}

The energy dissipation due viscous work $W$ is approximated by assuming vertical shear flows and the velocity profile in Eq. \ref{flat_velocity_profile}:

\begin{equation}
\label{flat_viscous_work}
W\approx \int_0^{t} \iiint_{V_\mathrm{lamella}} \mu \left( \frac{\partial U}{\partial z} \right)^2 \mathrm{d} \Omega \mathrm{d}t' = \int_0^t \frac{3\pi \mu \dot{h}_c^2 r_0^4}{2h^3}\ln \left( \frac{R}{r_0} \right) \mathrm{d}t'.
\end{equation}.

Using Eq. \ref{surface_engery}, \ref{flat_kinetic_energy}, \ref{flat_viscous_work}, and Young's relation $ \gamma \cos \theta = \gamma_{SG} - \gamma_{SL} $, the total energy balance reduces to

\begin{eqnarray}
\label{flat_energy_balance}
\gamma\left[2Rh+(1-\cos \theta)R^2+ (\delta+h)(\delta+h-4R_0) \right] \nonumber \\
+ \rho\left[ \frac{3\dot{h}_c^2 r_0^4}{10h}\ln\left( \frac{R}{r_0} \right) + \frac{2(\dot{h}_c^2-U_0^2)R_0^3}{3}-\frac{ \dot{h}_c^2V_{beyond}}{2\pi} \right] +\frac{W}{\pi}=0.
\end{eqnarray}

At the early stage, we can assume $\delta \ll R_0$, $\dot{h}_c \approx U_0$, $R\approx r_0$ as $R$ is small, so that $W$ becomes negligible and $\theta=\pi$. The volume conservation in Eq. \ref{flat_volume_conservation} results in $\delta \approx h$. With these assumptions and neglecting high order terms in $\delta$, Eq. \ref{flat_energy_balance} simplifies to
\begin{equation}
\label{flat_early_stage}
R^2+Rh-2R_0(\delta+h)= 0.
\end{equation}
A solution of Eq. \ref{flat_early_stage} can be approximated as $R \approx 2\sqrt{\delta R_0} \approx 2\sqrt{U_0R_0}\sqrt{t}$ at early stage, which is consistent with the experimental exponent reported in literature and in Bird, Tsai and Stone derivation \cite{BTS2009}.\

At the final stage, $\delta \approx 2R_0$, $ h \ll R $ and $\dot{h}_c \approx 0$, simplifying Equation \ref{flat_energy_balance} to
\begin{eqnarray}
\gamma[2Rh+(1-\cos \theta)R^2+h^2-4R_0^2] 
-\frac{2}{3}\rho U_0^2 R_0^3+ \frac{W}{\pi} =0,
\end{eqnarray}
Assuming $1-\cos\theta = \mathcal{O}(1)$, and approximating the viscous work $W$ with the expression used by Pasandideh-Fard et al. \cite{PQCM1996}, $W\approx \frac{\pi}{3}\rho U_0^2 (2R_0) (2R)^2/\sqrt{\mathrm{Re}}$, results in the same expression for the maximal spreading distance as the one found by Pasandideh-Fard et al., which was found to give a good approximation to the ratio
\begin{equation}
\label{PF_ratio}
\frac{R_\mathrm{max}}{R_0}=\sqrt{\frac{\mathrm{We}+12}{3(1-\cos \theta)+4 \frac{\mathrm{We}}{\sqrt{\mathrm{Re}}}}},
\end{equation}
where $R_\mathrm{max}$ is the spread distance at the final stage, which corresponds to the maximum spread distance.

\section{Appendix B: Case of a curved rigid surface}
\subsubsection{Geometry of the model}

Figure \ref{ExpSetup}b shows the schematic of a drop impacting a spherical ball. As in the previous section, we use rigid spheres with curvature radius $R_b$. Now, $R$ represents half of the total spreading length along a curved surface. Due to the change in geometry, $V_{beyond}$ can now be calculated as the intersection volume of two spheres of radius $R_b$ and $R_0$, and $V_\mathrm{beyond}+V_\mathrm{middle}$ as the intersection of two spheres of radius $R_b+h$ and $R_0$. $V_\mathrm{lamella}+V_\mathrm{middle}$ can be expressed as the intersection of two spherical sectors of radii and arc lengths $(R_b+h,\theta_1)$ and $(R_b,\theta_1)$, where $\theta_1=R/R_b$ is the arc length. As in the previous section, we assume that the original drop and the remaining part are spherical. Additionally, the height of the lamella is assumed to be much smaller than the surface radii, that is, $h\ll R_b$. 

\subsubsection{Flow in the lamella}
The lamella is assumes a semi-parabolic velocity profile, similar to the flar case. However, due to the curvature of the rigid surface, the radial velocity is modified as $U(s,z,t)=\overline{U}(t)\frac{s_0(t)}{s}\frac{z(2h-z)}{h^2}$ where $s$ is the curvilinear abscissa, defined as the arc length $s=R_b\theta$. The curvilinear abscissa spreading over time is $s_0(t)=R_b \arcsin(\frac{a}{R_b})$ where the half of the chord in $V_\mathrm{beyond}$ is
\begin{equation}
\label{curved_curvilinear_abscissa}
    a=\frac{\sqrt{(\delta+h)(\delta-h-2R_b)(\delta+h-2R_0)(2R_0+2R_b-\delta+h)}}{2(R_0+R_b-\delta)}.
\end{equation}
Therefore, the velocity profile of the lamella obtained by applying mass conservation is
\begin{equation}
\label{curved_velocity_profile}
U(s,z,t) = \frac{3 |\dot{h}_c(t)|s_0^2(t) }{4h^3(t)s}z(2h-z).
\end{equation}

\subsubsection{Governing equations}
The volume conservation is obtained by equating the virtual (beyond) volume with the lamella volume (in the curved case of Fig.\ref{ExpSetup}b). The volumes for the curved case account for the middle section of the impacting droplet. That is, the volume conservation is $V_\mathrm{middle+beyond} = V_\mathrm{lamella+middle}$. Using the geometric properties of the curved surface, these volumes are
\begin{eqnarray}
V_\mathrm{beyond} &=& \dfrac{\pi \delta^2 }{12(R_b +R_0 -\delta)}[(R_b +R_0 -\delta)(3R_b +3R_0 - \delta)-3(R_0 - R_b)^2], \\
V_\mathrm{middle+beyond} &=&   \dfrac{\pi (\delta+h)^2}{12(R_b +R_0 -\delta)} [(R_b +R_0 -\delta) (3R_b +3R_0+ 2h- \delta)-3(R_0 - R_b-h)^2], \\
V_\mathrm{lamella+middle}&=& \dfrac{2\pi h}{3} (1-\cos \theta_1)(3R_b^2+3R_b h+h^2),
\end{eqnarray}



where $\theta_1$ is the angle between the centerline and the edge of the lamella. In the limit of $R_b \rightarrow +\infty$, these volumes converge the flat surface case. By applying the volume conservation and simplifying it, it becomes
\begin{eqnarray}
\label{curved_volume_conservation}
(R_b+R_0-\delta) [ 8(1-\cos \theta_1)h(3R_b^2 +3R_b h +h^2) 
-(h+\delta)^2 (3R_b+3R_0+2h-\delta)] \nonumber \\
+3 (\delta+h)^2(R_0-R_b-h)^2=0.\
\end{eqnarray}

The total energy balance is computed similar to the flat case, but the curvature must be addressed in the computation of the surface and kinetic energies, and the viscous work. The expressions for the areas of the solid-liquid and liquid-gas interfaces are
\begin{eqnarray}
S_\mathrm{LS} &=& 2 \pi R_b^2(1- \cos \theta_1), \\
S_\mathrm{LG} &=& S_\mathrm{lamella}+S_\mathrm{remain},  \\
S_\mathrm{lamella} &=& 2 \pi (R_b+h) ^2 \left(\cos \theta_2- \cos \theta_1  \right) + 2\pi  h\left(R_b+ \dfrac{h}{2} \right) \sin \theta_1 ,\\
S_\mathrm{remain} &=& 4\pi R_0^2-2\pi R_{0}(\delta+h).
\end{eqnarray}
where $\theta_2$ is the angle between the centerline and the contact of a drop on a solid. The liquid-solid and liquid-gas interfaces are computed similar to the flat case. That is, the liquid-solid interface is computed as the surface area of a spherical cap, while the liquid-gas interface is computed as the sum of the surface areas of the lamella and the remaining droplet. The terms in the $S_\mathrm{lamella}$ equation represent the surface areas of the top and side of the lamella, respectively. $S_\mathrm{remain}$ is computed similar to the flat case. In the limit case when $R_b \rightarrow \infty $ \, and letting $R_b \theta_{1} \rightarrow R$ and $R_b \theta_{2} \rightarrow R_0^2-(R_0-\delta-h)^2 $, we see convergence to the terms $S_\mathrm{LS}$ and$S_\mathrm{LG}$ of the flat case.

Another distinction between the cases is captured in the radial component, $s(t)$, which now spans from $R_{b}\theta_2$ to $R_{b}\theta_1$. The surface energy is computed is using the curved surface areas, Eq. \ref{surface_engery}, and Young's relation. Calculations of the kinetic energy and viscous work are also done similar to the flat case. Therefore, the total energy balance for the curved case is
\begin{eqnarray}
\label{curved_energy_balance}
\gamma \left[2\pi (R_b+h)^2(\cos\theta_2-\cos\theta_1) + 4\pi R_0^2-2\pi R_0(\delta+h) + 2\pi h ( R_b+ \dfrac{h}{2} ) \sin\theta_1  -2\pi R_b^2\cos\theta (1-\cos\theta_1)\right]  \nonumber \\
-4\gamma \pi R_0^2  + \pi \rho\dot{h}_c^2 \left[ \dfrac{2 R_{0}^3}{3} -\dfrac{V_\mathrm{beyond}}{2\pi} + \dfrac{3 s_0^4}{10 h} \ln\left(\dfrac{\theta_1}{\theta_2}\right)\right] -\dfrac{2\pi \rho U_0^2 R_0^3 }{3} + W=0,
\end{eqnarray}
where 
\begin{equation}
\label{curved_viscous_work}
W=\int_0^t \frac{3\pi \mu \dot{h}_c^2 s_0^4}{2h^3}\ln\left(\frac{\theta_1(t)}{\theta_2(t)}\right) dt'.
\end{equation}




At the early stage, we can assume $R \approx s_0 \ll R_0$, $R \ll R_b$ and $\delta \ll R_0$, $\delta \ll R_b$ so that $\sin \theta_1 \approx R/R_b$ and negligible viscous work. Simplifying volume conservation and total energy balance give
\begin{equation}
R^2 \approx \frac{(\delta+h) R_b R_0}{R_b+R_0}, \\
\end{equation}
and
\begin{equation}
    R \approx \frac{R_0(\delta+h)}{h}.
\end{equation}
Combining these equations gives and letting $\delta \approx U_0 t$ at the early stage gives the approximation
\beq
R^2 = \delta \left(1 + \frac{1}{1-R/R_0} \right) \frac{R_b R_0}{R_b + R_0} \,.
\eeq 
If we assume $R/R_0 \ll 1$ for the early time of spreading, then 
\begin{eqnarray}
&& R^2 \approx 2 \delta \frac{R_b R_0}{R_b + R_0} =  2 \frac{R_b R_0}{R_b + R_0} U_0 t = 4 \frac{R_b R_0^2}{R_b + R_0} \frac{U_0 t}{2 R_0}\\
&& R(t) \approx R_0\left( \frac{4R_b}{R_b+R_0} \right)^{1/2} \left( \frac{U_{0}}{2R_{0}} t \right)^{1/2}  
\end{eqnarray}
This implies that during the early stage, a droplet would spread farther on a curved surface than on a flat one.

At the later stage, assuming $\delta \approx 2R_0$, $\theta_2 \ll 1$, $\dot{h}_c\approx 0$, $h \ll R_b$, the total energy balance can be expressed as
\beq
\mathrm{We} \left[ 1-\frac{4}{\sqrt{\mathrm{Re}}} \frac{R(t)^2}{R_0^2} \right]=6 \frac{R_b^2}{R_0^2} (1-\cos \theta_{\mathrm{Advance}})(1-\cos \theta_1) - 12 \,. \label{sp_dyn_curve}
\eeq
In the limit of the flat case ($R_b \rightarrow +\infty$ and $1-\cos \theta_1 \approx \frac{1}{2}(R^2(t)/R_b^2)$), Equation \ref{curved_energy_balance} reduces the drop-spreading equation of Pasandideh-Fard et al. \cite{PQCM1996} in Eq. \ref{PF_ratio}.

\section*{Acknowledgements}
 This work was supported by the National Science Foundation Grant No. ISO-2120739.

\section*{Data availability}
The data used in this study are deposited in the public repository (DOI 10.17605/OSF.IO/WX3ET; https://osf.io/wx3et/). 

\section*{Author contributions}
M.L. and S.J. conceived the idea. M.L. and J.H. performed experiments. M.L., J.H., and S.J. developed a theoretical model. M.L. wrote the original draft. All authors revised the manuscript and contributed to the final version. 

\section*{Competing interests}
The authors declare no competing interests.

\newpage

\section{References}
\bibliographystyle{apsrev4-1}

\end{document}